\documentclass[11pt]{article}
\usepackage{amsfonts,bm,bbm,amssymb,caption}
\usepackage{upgreek} % per fare lettere greche con font diverso  $\upmu$
\usepackage{mathrsfs}
\usepackage{amsmath}	
\usepackage{cite}
\usepackage{graphicx}
\usepackage{rotating}
\usepackage{hyperref}
\usepackage{verbatim}
\usepackage[all]{xy}

\csname @addtoreset\endcsname{equation}{section}
\newcommand{\beq}{\begin{equation}}
\newcommand{\eeq}{\end{equation}}
\newcommand{\bea}{\begin{eqnarray}}
\newcommand{\eea}{\end{eqnarray}}
\newcommand{\ba}{\begin{array}}
\newcommand{\ea}{\end{array}}
\newcommand{\bit}{\begin{itemize}}
\newcommand{\eit}{\end{itemize}}
\newcommand{\nn}{\nonumber}

\newcommand{\complesso}{{\ \hbox{{\rm I}\kern-.6em\hbox{\bf C}}}}
\newcommand{\reale}{{\hbox{{\rm I}\kern-.2em\hbox{\rm R}}}}
\newcommand{\p}{\partial}
\renewcommand{\a}{\alpha}
\renewcommand{\b}{\beta}
\newcommand{\g}{\gamma}

\renewcommand{\d}{\delta}

\newcommand{\e}{\epsilon}
\newcommand{\Er}{{\mathcal{E}}}

\renewcommand{\k}{\kappa}
\renewcommand{\l}{\lambda}
\renewcommand{\L}{\Lambda}

\newcommand{\m}{\mu}

\newcommand{\n}{\nu}
\renewcommand{\r}{\rho}
\newcommand{\s}{\sigma}

\renewcommand{\S}{\Sigma}

\renewcommand{\t}{\theta}

\newcommand{\x}{\xi}

\newcommand{\om}{\omega}
\newcommand{\Om}{\Omega}

\begin{document}

\begin{titlepage}
\begin{flushright}
%hep-th/??????\\
CECS-PHY-13/05
\end{flushright}
\vspace{2.5cm}
\begin{center}
\renewcommand{\thefootnote}{\fnsymbol{footnote}}
{\LARGE \bf C-metric with a conformally coupled scalar field}
\vskip 4mm
{\LARGE \bf in a magnetic universe}
\vskip 30mm
{\large {Marco Astorino\footnote{marco.astorino@gmail.com}}}\\
\renewcommand{\thefootnote}{\arabic{footnote}}
\setcounter{footnote}{0}
\vskip 10mm
{\small \textit{
Centro de Estudios Cient\'{\i}ficos (CECs), Valdivia,\\ 
Chile\\}
}
\end{center}
\vspace{5.2 cm}
\begin{center}
{\bf Abstract}
\end{center}
{In Einstein-Maxwell gravity with a conformally coupled scalar field, the black hole found by Bocharova, Bronnikov, Melnikov, and Bekenstein breaks when embedded in the external magnetic field of the Melvin universe. The situation improves in the presence of acceleration, allowing one to build a magnetised and accelerating BBMB black hole with a thin membrane. But to overcome this and others disadvantages of BBMB spacetimes, a new class of black holes, including the rotating case, is proposed for the conformal matter coupling under consideration.}
\end{titlepage}

\tableofcontents

\section{Introduction}
%\label{intro}

Solution generating techniques are a very powerful tool in general relativity. Taking advantage of the integrability properties of the system and its symmetries, they are not only a mere mechanism to build solutions hardly directly integrable from the (non-linear system of partial differential) equations of motion, but their formalism is also useful to deepen conceptual problems in gravity, such as the Geroch conjecture\footnote{The Geroch conjecture was proven by Hauser and Ernst in \cite{ernst0}. It states that any axially symmetric electro-vacuum spacetime can be generated from the Minkowski one by Kinnersley-Chitre transformations.} or black hole uniqueness. \\
Recently the Ernst solution generating technique, originally developed for axisymmetric spacetimes in Einstein general relativity without a cosmological constant \cite{ernst1}, possibly coupled with Maxwell electromagnetism \cite{ernst2}, was extended to the presence of a minimally or a conformally coupled scalar field in \cite{marcoa}. The latter theory admits a black hole discovered by Bocharova, Bronnikov, Melnikov and Bekenstein (BBMB henceforward) in \cite{BBM} and \cite{bekenstein1}, \cite{bekenstein2}. This was the first counterexample to the no-hair conjecture for black holes. Thanks to the generalised Ernst methods it was possible to extend the Harrison transformation, which allows one to embed an asymptotically flat and axisymmetric spacetimes in the Melvin magnetic universe. So the family of magnetised black hole, known as Ernst solutions, were widened to enclose also the BBMB black hole \cite{marcoa}. The presence of the scalar field, which for the BBMB black hole is divergent on the event 
horizon, makes 
the black hole break when immersed in an external magnetic field, that is the magnetised solution displays curvature singularities on some points of the horizon. In the presence of the cosmological constant the divergence of the scalar field can be neutralised because it is hidden behind the event horizon, but unfortunately neither a solution generating technique nor a Harrison transformation is available at the moment for this system. Some attempts to adapt the Ernst method to the presence of the cosmological constant were done in \cite{charmousis} and \cite{ernst-lambda}, small progresses were achieved there (for instance the generalisation of the Melvin magnetic universe in presence of the cosmological constant), but basically the problem still remains open.\\
Recently also a C-metric was discovered, in \cite{greco} and \cite{andres-hideki}, for Einstein-Maxwell theory with a conformally coupled scalar field, which is interpreted as a pair of accelerating BBMB black holes. A typical feature of these accelerating solutions is that the acceleration is provided by a conical singularity, physically interpreted as a string or a strut pulling or pushing respectively the two black holes. Ernst, in \cite{ernst-remove}, using a Harrison transformation has shown how to regularize these accelerating (when intrinsically charged) solutions, removing the deficit or excess angle of the conical singularity by embedding the C-metric (in the case without the scalar field) in an external magnetic field. Actually this regularisation mechanism was invoked in \cite{greco} but the Harrison transformation in presence of a scalar field was not known, because the solution generating technique \cite{marcoa} was not available at that time. Furthermore it is worth to note that the 
accelerating BBMB solution has a better behaved scalar field than the non accelerating one, because the scalar field blows up only on one pole of the event horizon and not on the whole surface. Similarly to the case with cosmological constant (almost all) of the divergences are hidden inside the event horizon. \\
Since now we are in possession of the technology able to magnetised spacetimes in the Einstein-Maxwell theory with a conformally coupled scalar field, it would be interesting to explore the possibility of regularizing the c-metric with a scalar hair by embedding it in an external magnetic field, this point is addressed in section \ref{Acc-BBMB-Melvin}. And since the accelerating BBMB has a more regular scalar field with respect to its static version, we hope also to be able to remove the naked singularities present in the magnetised BBMB black hole. \\
Furthermore one virtue of the Ernst solution generating technique, which at the beginning was probably the main motivation for its discovery, is to generate rotating solutions starting from a static seed, for instance obtaining the Kerr spacetime from the Schwarzschild black hole. So it is natural, with the help of the generalised solution technique, to explore the possibility to generate a rotating version of the BBMB black hole, which is still unknown. Unfortunately the standard methods that work for the case without the scalar hair fails, so in section \ref{const-scalar-hairy-bh} a rotating, scalar hairy black hole is considered to overcome this and other disadvantages typical of BBMB spacetimes.\\ 
While the existence of a (minimal or) conformally coupled scalar field is not proven in gravitational physics, they are theoretically widly used especially in cosmology for studying dark energy and dark matter. On the other hand the astrophysical interest in black holes embedded in an external magnetic source, such as the Melvin universe, comes from the fact that currents in the accretion disk around a massive black hole, especially the ones at the center of galaxies, can presumably generate such kind of magnetic fields. \\

%%%%%%%%%%%%%%%%%%%%%%%%%%%%%%%%%%%%%%%%%%%%%%%%%%%%%%%%%%%%%%%%%%%%%%%%%%%%%%

\section{Accelerating BBMB Black Hole in Melvin magnetic universe}
\label{Acc-BBMB-Melvin}

Thanks to the solution generating techniques developed in \cite{marcoa} for Einstein-Maxwell gravity theory with a conformally (and minimally) coupled scalar field, it is now possible to embed the accelerating, scalar hairy black hole discovered in \cite{greco} in the Melvin magnetic universe. 

\subsection{C-metric with a conformal scalar hair}
\label{conformal-C-metric}

Consider the action for general relativity coupled to the Maxwell electromagnetic field and to a conformally coupled  self  interacting scalar field $\Psi$:
\beq  \label{minimal-action}
                       I[g_{\m\n}, A_\m, \Psi] =  \frac{1}{16 \pi G}  \int d^4x  \sqrt{-g} \left[ \textrm{R} - F_{\m\n} F^{\m\n} - \k \ \left( \nabla_\m \Psi \nabla^\m \Psi  + \frac{ \textrm{R}}{6} \Psi^2   \right) \right] \ \ \ .
\eeq

The gravitational, electromagnetic and scalar field equations are obtained by extremising with respect to metric $g_{\m\n}$, the electromagnetic potential $A_\m$ and the scalar field $\Psi$ respectively
\bea  \label{min-field-eq}
                        &&   \textrm{R}_{\m\n} -   \frac{\textrm{R}}{2}  g_{\m\n} =  \k \left(  T^{(EM)}_{\m\n} +  T^{(S)}_{\m\n} \right)  \quad ,       \\
                        &&   \partial_\m ( \sqrt{-g} F^{\m\n}) = 0  \ \quad , \\
      \label{scalar-eq} &&   \Box \Psi=\frac{1}{6} \textrm{R} \Psi \quad ,
\eea
where
\bea
      T^{(EM)}_{\m\n} &=&   \frac{1}{4 \pi \m_0} \left( F_{\m\r}F_\n^{\ \r} - \frac{1}{4} g_{\m\n} F_{\r\s} F^{\r\s} \right)~, \\
      T^{(S)}_{\m\n} &=&  \p_\mu\Psi\p_\nu\Psi-\frac{1}{2}g_{\mu\nu} \p_\s\Psi\p^\s\Psi+\frac{1}{6}\left[g_{\mu\nu}\square-\nabla_\mu\nabla_\nu+G_{\mu\nu}\right]\Psi^2~ .   \label{eltr-eq}
\eea

In this section we are interested in static and axisymmetric space-times characterized by two commuting killing vectors described by the Weyl metric
\beq \label{weyl-metric}
                         ds^2 =  -f d\varphi^2 + f^{-1} \left[ R^2 dt^2 - e^{2\gamma}  \left( d R^2 + d z^2 \right) \right] \ ,
\eeq
where the functions $f,\gamma$ depend only on the coordinates $(R,z)$ and $\k=8\pi G$, while the electromagnetic potential will be taken of the form $ A = A_t(R,z) dt + A_\varphi(R,z) d\varphi $\footnote{It is shown by Carter in \cite{carter} (theorem 7) that this is the most generic circular electromagnetic field, compatible with the circular metric (\ref{weyl-metric}).}. An accelerating black hole solution for this model was found in \cite{greco} (see also \cite{andres-hideki}), for null cosmological constant and electromagnetic charge, it is
\bea \label{greco-metric}
     && ds^2=\frac{1}{\left(1+A r\cos{\theta}\right)^2}\left[ -\frac{Q(r)}{r^2} dt^2+\frac{r^2}{Q(r)} dr^2+\frac{r^2}{P(\theta)} d\theta^2+r^2\sin^2{\theta} P(\theta) d\varphi^2 \right] \\
     && Q(r) = \left(1-A^2r^2\right)\left(r-m\right)\left(r-\frac{m}{1+2Am}\right)~, \label{Q(r)}\\
     && P(\theta) =   \left(1+A m\cos{\theta}\right)\left(1+\frac{Am}{1+2Am} \cos{\theta} \right)~, \\
     && \Psi(r ,\theta) = \sqrt{\frac{6}{\k}}\:\frac{m(Ar\cos{\theta}+1)}{r(1+Am)+m(Ar\cos{\theta}-1)}~. \\
\eea
$A$ and $m$ represent respectively the acceleration and the mass parameters of the black hole, we will consider them positive. Actually this metric (\ref{greco-metric}) is interpreted as a pair of black holes  with a conformally coupled scalar hair uniformly accelerating apart along the axis $\theta=0$. The inner $r_-$, outer $r_+$ and accelerating $r_A$ horizons are given by
\beq
      r_-= \frac{m}{1+2Am} \qquad , \quad \qquad r_+ = m \qquad , \quad \qquad r_A=\frac{1}{A}
\eeq
In order for the roots of the polynomial $Q(r)$ in (\ref{Q(r)}) to be ordered according to C-metric interpretation, the parameters have to satisfy the following relation:
\beq   \label{para-range}
       0 \le A r_- \le A r_+ \le 1 \quad .
\eeq
Moreover, as explained in \cite{interpret-c}, C-metrics usually have a hidden parameter $C$ in the range of azimuthal coordinate $\varphi \in (-C\pi, C\pi]$ . When the acceleration parameter $A$ goes to zero the black hole found by Bocharova, Bronnikov and Melnikov in \cite{BBM}, and then studied by Bekenstein in \cite{bekenstein1} - \cite{bekenstein2}, is recovered. %Henceforth we will call this $A=0$ metric BBMB.
In that case ($A=0$) there is no accelerating horizon and both the inner and outer horizon coincide: $r_\pm=m$.\\
It is worth to examine the regularity of the axis of symmetry because in the literature this point is often not clear. This ambiguousness usually arises from a different choice of the radial coordinate $r$. As pointed out in \cite{hong-teo}, our radial coordinate choice is motivated by the facts that $(i)$ the no accelerating limits are more clear, $(ii)$ when the C-metric is rotating there are no torsion singularities (that is rotating conical singularities, which generates closed time-like curves\footnote{This feature makes the two coordinate choices not physically equivalent in presence of rotation.}), $(iii)$ moreover the interpretation of the extremal case is comprised in the standard case (the string and acceleration is not disappearing in the extremal case), $(iv)$ and finally the simpler algebra makes the position of the horizon clearer.  To study the conicity of the metric (\ref{greco-metric}) we consider a small circle around the half-axis $\t = 0$ (with $t, r$ constant). For the above range of $\
varphi$, we
obtain
\beq
{\hbox{circumference}\over\hbox{radius}} 
 =\lim_{\t \to 0} {2\pi C P(\t) \sin \t \over \t} = 2 \pi C \left(1 + A m + \frac{Am}{1+2Am}+\frac{A^2m^2}{1+2Am} \right)\;.
\eeq
When this value is different from $2\pi$, the metric (\ref{greco-metric}) has at least a conical singularity in $\theta=0$. Similarly, around the other half axis $\t=\pi$, we have 
\beq
{\hbox{circumference}\over\hbox{radius}} 
 =\lim_{\t \to \pi} {2\pi C P(\t) \sin \t \over \pi - \t} = 2 \pi C \left(1 - A m - \frac{Am}{1+2Am}+\frac{A^2m^2}{1+2Am} \right)\;.
\eeq
A deficit angle is interpreted as a semi infinite cosmic string pulling the BBMB black hole along the half axis with a force proportional to the tension of the string (i.e. a $T_{\m\n}$ localised on the string and proportional to the deficit angle), conversely an excess angle is interpreted as a strut pushing the black hole.\\
Because the conicity of the conical singularities are different on the two half axes, in general it is not possible to remove them simultaneously, fixing the value of the constant $C$. 
Henceforward, to avoid the conical singularity for $\t=0$, we will set\footnote{Alternatively a new axial angular coordinate, with canonical period $2\pi$,  can be defined dilatating the old one by a factor $C^{-1}$.}
\beq
       C = \left(1 + A m + \frac{Am}{1+2Am}+\frac{A^2m^2}{1+2Am} \right)^{-1}\;.
\eeq
One can not even remove  the second conical singularity by a non trivial fine-tuning between the parameters such that $ A m + \frac{Am}{1+2Am} =0$
because, apart from the trivial cases for  $A=0$ or $m=0$, corresponding to Schwarzschild or Minkowski space-times in accelerating coordinates respectively, the only remaining possibility is $Am=-1$; but unfortunately it is outside the range of permitted parameters (\ref{para-range}). 
Note that the rotating solution of \cite{andres-hideki} lacks of conical singularity, though it is accelerating, because it does not have a proper mass term\footnote{As can be seen from the vanishing acceleration limit.}. \\
Usually, as was found by Ernst himself in \cite{ernst-remove} for the case of vanishing scalar field, it is possible to introduce an external magnetic field to remove this residual conical singularity from the charged C-metric. We will do the same with a non null scalar field in the subsection \ref{remove-cony}. \\
We finally observe that, although the non-accelerating case of solution (\ref{greco-metric}) has a divergent scalar field $\Psi(r,\t)$ behaviour on the whole outer horizon $r=m$,  when $A$ is non null the scalar field is well behaved except on one pole $(r=r_+,\t=\pi)$:
  \beq
        \Psi(r_+,\t) = \frac{Am\cos \t +1}{Am(1+\cos \t)}\;~,
  \eeq
where it is divergent. The scalar field divergences were the origin of the problems in the magnetised BBMB solution in \cite{marcoa}, thus a better behaved scalar field on the horizon is favourable for magnetising purposes. \\

\subsection{C-metric with a conformal scalar hair in the Melvin magnetic universe}
\label{magnetic-bbmb}

Here we want to embed the metric (\ref{greco-metric}), which we will consider our seed, in the external magnetic field of the Melvin magnetic universe. To do that it is necessary to have the Harrison transformation for the theory under consideration. Using the results of \cite{marcoa} we can write\footnote{One has just to pass to the Einstein frame with a conformal transformation, apply the desired transformation (in this case the Harrison one) and afterwards come back to the Jordan frame.} such kind of magnetising transformation in the solution space of the Einstein-Maxwell theory of gravity with a conformally coupled scalar field.  In terms of the Ernst potentials, for uncharged\footnote{A Harrison transformation preserving staticity is generalised in section \ref{app-acc-cost-bh} for a particular kind of charged seed.} and static seed space-times, the Harrison transformation is given by
\beq	\label{harrison-jordan}																	 	  
       \Er_0 \longrightarrow \Er = \frac{\Er_0 - \frac{B^2}{4} (1 -\frac{\k}{6} \Psi^2)^2 \Er_0^2}{\left[1-\frac{B^2}{4} \left(1 -\frac{\k}{6} \Psi^2
        \right) \Er_0\right]^2} \qquad \quad , \qquad \quad  \mathbf{\Phi_0} \longrightarrow  \mathbf{\Phi} = \frac{\frac{B}{2}\left(1 -\frac{\k}{6} \Psi^2 \right)\Er_0}{1-\frac{B^2}{4} \left( 1 -\frac{\k}{6} \Psi^2 \right) \Er_0} \quad .
\eeq
We recall the definition of Ernst complex potentials which, just for this particular uncharged and static seed case, remain real:
\beq
       \Er := f - \mathbf{\Phi}^2    \qquad , \qquad    \mathbf{\Phi}:=A_\varphi \quad . 
\eeq 
The Ernst potentials for the seed metric (\ref{greco-metric}) are obtained comparing it with the Weyl one (\ref{weyl-metric}):
\beq
       \mathbf{\Phi}_0 = 0   \qquad , \qquad  \Er_0 = f_0 = -  \frac{P(\theta)~ r^2 \sin^2{\theta}}{\left(1+A r\cos{\theta}\right)^2}~.
\eeq 
So, while the function $\g(r,\t)$ remains unchanged as in the Weyl metric (\ref{weyl-metric}), the magnetised $f$ is given by 
\beq \label{harry-transf}
f= \Er + \mathbf{\Phi}^2 = \frac{f_0}{\Lambda^2(r,\theta)}  \qquad  \textrm{where} \qquad \Lambda(r,\theta)= 1-\frac{B^2}{4} \left(1 -\frac{\k}{6} \Psi^2 \right) f_0 \quad .
\eeq
Finally the resulting magnetised version of the accelerating C-metric (\ref{greco-metric}) becomes
\beq \label{magn-greco-metric}
      ds^2=\frac{\Lambda^2(r,\t)}{\left(1+A r\cos{\theta}\right)^2} \left[ -\frac{Q(r)}{r^2} dt^2+\frac{r^2}{Q(r)} dr^2 + \frac{r^2}{P(\theta)} d\theta^2 + \frac{r^2\sin^2{\theta} P(\theta)}{\Lambda^4(r,\t)} d\varphi^2  \right]   \ \ ,
\eeq
supported by the magnetic field
\beq \label{unch-c-acc}
        A_\varphi =  \frac{\frac{B}{2}\left(1 -\frac{\k}{6} \Psi^2 \right)f_0}{1-\frac{B^2}{4} \left( 1 -\frac{\k}{6} \Psi^2 \right) f_0}  \quad .
\eeq
The conical singularity present at the point $\t=\pi$ can not be removed by the addition of an external magnetic field because, as shown in \cite{marcoa}, the excess or deficit angle which stems from embedding conformal scalar hairy black holes in the Melvin universe is proportional to both the intensity of the external magnetic field $B$ and the "intrinsic" electromagnetic charge $e$ of the seed black hole. Since in this case the seed solution (\ref{greco-metric}) is eletromagnetically neutral, it is not possible to do a fine tuning between the parameters ($A,m,e,B$) to elide the nodal singularity, exactly as in the case of null scalar field \cite{ernst-remove}. In the next section \ref{remove-cony}, a intrinsically charged solution will be considered, so that we will be able to apply this Ernst trick. \\  
The scalar curvature invariants of this metric (\ref{unch-c-acc}), such as R$^{\m\n}$R$ _{\m \n}$ and R$^{\m\n\s\l}$R$_{\m\n\s\l}$, are divergent only on the pole  $(r=r_+ ,\  \t=\pi)$. As expected this is a reminiscence of the singular character of the field $\Psi(r,\t)$ on that pole. So the magnetised C-metric solution is slightly better behaved than the non accelerating one of \cite{marcoa}, but still it remains singular. Nevertheless now it can be used, via the cut and paste procedure of \cite{empa-brane}, to build a regular black hole on the brane; this will be done in section \ref{thin}. \\

\subsection{Removing the conical singularity from the scalar hairy, charged C-metric}
\label{remove-cony}

The Ernst method \cite{ernst-remove} to remove the conical singularity typical of the C-metric space-time consists of embedding it in an external magnetic field. In order to achieve that it is essential to have an interaction between the intrinsic charge of the black hole (which could be of electric or magnetic type) and the external field (which either can be electric). The simpler implementation remains within a static framework, so it is necessary that intrinsic electric charge and external charge are of the same type. For instance we will consider an intrinsically magnetically charged accelerating BBMB black hole embedded in an external magnetic field. It is easy to see, via the electromagnetic duality in four dimensions, that the same result can be obtained by an electrically charged accelerating black hole embedded in an external electric field.  On the other hand when the intrinsic and external electromagnetic charges are of a different type the metric becomes stationary due to the appearance of a $\
\overrightarrow{E} \times \overrightarrow{B}$ circulating momentum flux in the stress-energy tensor, which serves as a source for a twist potential. To be more precise in order to preserve the staticity of the seed space-time, even when is not electromagnetically neutral, the Ernst potential $\Er$ must remain real.  \\
Thus let's consider as a seed metric an accelerating BBMB black hole with intrinsic magnetic charge $g$. It has the same form as the uncharged one (\ref{greco-metric}), but the matter fields are
\beq \label{matter-acc}
     A_\varphi=-g \cos(\t) \qquad , \qquad \qquad \Psi(r,\t) = \pm \sqrt{\frac{6}{\k}}  \frac{\sqrt{m^2-g^2(1+2Am)} \:(Ar\cos{\theta}+1)}{r(1+Am)+m(Ar\cos{\theta}-1)} \ .
\eeq
Because the metric is charged we cannot use the Harrison transformation directly in the conformal frame, but we have to shift it in the minimal frame, magnetise the shifted metric and then come back in the Jordan frame, as explained in \cite{marcoa}. The resulting magnetised metric remains formally the same as the uncharged case (\ref{magn-greco-metric}), also the scalar field remains the same as (\ref{matter-acc}), but the magnetic potential becomes
\beq
      A_\varphi = - \frac{ g \cos \t + \frac{B}{2} g^2 \cos^2 \t + \frac{B}{2} \left\{ 1 - [m^2-g^2(1+2mA)] \left[\frac{1+Ar\cos \t}{(1+Am)r+m(Ar\cos \t -1 )} \right]^2  \right\} \frac{P(\t) r^2 \sin^2 \t}{(1+Ar\cos\t)^2}  }{\Lambda(r,\t)} \nn
\eeq
and $\L$, for the charged case, modifies in
\bea
     \Lambda(r,\t) &=& 1 + g B \cos \t + \frac{g^2 B^2}{4} \cos^2 \t + \nn \\ &+&   \frac{B^2}{4} \left\lbrace 1 - [m^2-g^2(1+2mA)] \left[\frac{1+Ar\cos \t}{(1+Am)r+m(Ar\cos \t -1 )} \right]^2   \right\rbrace   \frac{P(\t) r^2 \sin^2 \t}{(1+Ar\cos\t)^2}  \ \    .     \nn 
\eea
Thus the introduction of the additional  parameter $B$ related to the external electromagnetic field makes possible the removal of the conical singularity from both the north and south poles. Expanding the metric around $\t=0$, as done in section \ref{conformal-C-metric}, it is possible to pull out the angular deficit or excess in $\t=0$ by just rescaling the $\varphi$ coordinate
\beq
      \varphi \leadsto \phi=\varphi \frac{(1+Am)[1+Am/(1+2Am)]}{\left( 1 + g B/2 \right)^4}  \ \ .
\eeq
In order to eliminate also the conical singularity from $\t=\pi$ one has to fix a particular relation between the parameters $A,m,B,g$:
\beq \label{regul-const1}
       m A = \frac{g B ( 4 + g^2 B^2)}{4- 4 g B + 6 g^2 B^2 - g^3 B^3 + g^4 B^4/4} \ \ .
\eeq  
From a physical point of view the removal of the conical singularity corresponds to removing the string (or strut) in charge to provide the acceleration to the C-metric. It means that the acceleration of the black hole pair is entirely provided by the interaction force between the intrinsic electromagnetic charge of the black hole and the external magnetic field.\\
%In the non-relativistic limit (??  controllare ??), that is 
For small values of the electromagnetic field, $gB<<1$, the latter equation coincides with the Newtonian force felt by a massive magnetic monopole, of intensity $g$, in a uniform magnetic field whose strength is proportional to $B$ (or alternatively, via electromagnetic duality, the weak electric field limit corresponds to an electric charge in a uniform electric field)
$$m A \approx g B \ \ .$$ 
In fact this represents the non-relativistic limit, i.e. $A<<1$, as can be seen by inverting (\ref{regul-const1}) and expanding for small acceleration parameter $A$:
$$   gB=2 \ \frac{(\frac{1+3mA}{1-mA})^{1/4}-1}{(\frac{1+3mA}{1-mA})^{1/4}+1} \  \approx \  mA \qquad    $$ 
Usually these accelerating metrics, once regularised with the Ernst procedure, are of a certain interest because they provide a description of pair production of black holes in a magnetic field, as first pointed out in \cite{kastor} (see also \cite{hawking}). Unfortunately this picture in the context of BBMB black hole is ruined. In fact, despite of the removal of the conical singularities and the strut/string interpretation related to that, not even the addition of the electromagnetic charge to the accelerating hairy metric is sufficient to make it regular, because of the presence of a curvature singularity on the pole $(r=m,\t=\pi)$, hence the presence of a singularity not hidden behind an event horizon. This is due to the divergence of the scalar field, of the seed metric, at that point.
\\
\newpage

\subsection{Thin shell regularisation: \\Magnetised and accelerating BBMB black hole on the brane}
\label{thin}

It is possible to regularise at the same time both the conical singularity and the curvature singularity of the metric (\ref{magn-greco-metric}) localised at $\t=\pi$. We will take advantage of the same procedure used in \cite{empa-brane} to remove the conical singularity from the uncharged C-metric. The basic idea is to consider the regular half part of the solution (\ref{magn-greco-metric}), that is the one with $0\le \t\le \pi/2$,  to cut away the resting part for $\pi/2 \le \t\le \pi$ and then gluing into another copy of the regular one. While the continuity of the metric is assured, the price to pay is the introduction of an extra energy momentum tensor term $T^\S_{\m\n}=\d(\frac{\pi}{2}-\t) S_{ij} e^i_\m e^j_\n$, localised on the $\bar{\t}=\pi/2$ surface $\S$, to regularise the discontinuity of the first fundamental form on the pasting surface $\t=\pi/2$. Generalised junction conditions, for the theory we are considering, were discussed in \cite{maeda-k}; the thin shell of extra matter content can be 
quantified as follows. First let's define $h_{ij}$, the metric on the three-surface characterised by constant $\t$, and the  normalised outward orthogonal vector to the three-surface
\beq
      n^\m = \left[0 \ ,\ 0 \ ,\ \frac{\sqrt{P(\t)}(1+A r \cos \t)}{r \Lambda(r,\t)} ,\ 0 \ \right]
\eeq
So the extrinsic curvature on the three-surface is given by: 
\beq
      K_{ij} = \nabla_i n_j = \frac{\sqrt{g(\t)}(1+Ar\cos \t)}{2 r \L(r,\t)} \frac{ d}{d\t}  h_{ij}
\eeq
The induced surface stress energy tensor is given by
\bea
     S^{(S)}_{ij} &=& -\frac{1}{8\pi G} \left\{ \Big[K_{ij}\Big]^{\bar{\t}+}_{\bar{\t}-} \left(1-\frac{\k}{6} \Psi^2 \right) - h_{ij} \Big[K\Big]^{\bar{\t}+}_{\bar{\t}-} \left(1-\frac{\k}{18} \Psi^2 \right) \right\} = \\
      &=& \frac{A \ h_{ij}}{2\pi G \ \L(r,\bar{\t})} \begin{bmatrix}
   1-\frac{m(1+Am)}{2r(1+2Am)} & 0 & 0 \\
      0 & 1-\frac{m(1+Am)}{2r(1+2Am)} & 0 \\
   0 & 0 &  1+ \frac{\p_\t \log\L(r,\t)\big|_{\bar{\t}}}{Ar} \ \end{bmatrix} \nn  \\
   &+&\frac{A \Psi^2 \ h_{ij}}{36\pi G \ \L(r,\bar{\t})} \begin{bmatrix}
   \frac{\p_\t \log\L(r,\t)\big|_{\bar{\t}}}{Ar}+\frac{m(1+Am)}{2r(1+2Am)} & 0 & 0 \\
      0 & \frac{\p_\t \log\L(r,\t)\big|_{\bar{\t}}}{Ar}+\frac{m(1+Am)}{2r(1+2Am)} & 0 \\
   0 & 0 &  \frac{\p_\t \log\L(r,\t)\big|_{\bar{\t}}}{-2Ar}-\frac{m(1+Am)}{r(1+2Am)} \ \end{bmatrix} \ . \nn
\eea
Eventually also the electromagnetic field contribution may be taken into account, in the usual way:
$$ S^{(EM)}_{\ ij} = \lim_{\bar{\t}- \t = \epsilon \rightarrow 0} \int_{-\epsilon}^{+\epsilon} T^{(EM)}_{ij} dn$$
\\
%%%%%%%%%%%%%%%%%%%%%%%%%%%%%%%%%%%%%%%%%%%%%%%%%%%%%%%%%%%%%%%%%%%%%%%%%%%%%%

\section{Black holes with a conformally coupled constant scalar field}
\label{const-scalar-hairy-bh}

As we have seen in the previous sections or as it is known from the literature the BBMB solution, the actual model of a black hole with a conformally coupled scalar field reveals several drawback or disadvantages, which are not present in ordinary black holes of the Kerr-Newman family, let's list some:
\begin{itemize}
\item[$i)$] The scalar field is divergent on the horizon. 
\item[$ii)$] The space-time is unstable under linear perturbations \cite{Bronnikov}.
\item[$iii)$] When embedded in a external magnetic field it breaks down: it discloses curvature singularities on the horizon \cite{marcoa}.
\item[$iv)$] The introduction of the cosmological constant can hide the whole scalar field singularity behind the horizon, while the introduction of the acceleration cures just some divergences, but not all. These residual scalar field singularities, not hidden inside the event horizon, often causes naked singularities in the solution generating process, as seen in the previous section \ref{magnetic-bbmb} and in \cite{marcoa}.\footnote{To be more precise, due to the scalar field divergence, some of the $SU(2,1)$ Kinnersley symmetry transformations, studied in \cite{marcoa}, involves unbounded quantities when applied to the BBMB metric in the conformal frame. In this sense BBMB black hole is not a "physically good" seed for the solution generating technique.}  
\item[$v)$] The BBMB black hole carries just a dichotomic, secondary hair, in the sense that there is not a continuous parameter associated to this scalar hair. There is no non-extremal extension, which might make the extra parameter continuous \cite{zannias}. Moreover due to the extremality its entropy is null.
\item[$vi)$] It  does not have a continuous limit to the Schwarzschild or Reissner-Nordstrom black hole. In \cite{marcoa} it is shown how, from a generalisation of the the BBMB solution, the Penney one, in the conformal frame it is possible to reach the  Reissner-Nordstrom and the Schwarzschild black hole. But this is not an admissible physical process because, in order to do that, one has to pass through naked singularities. 
\item[$vii)$] A stationary version of the BBMB black hole is not known.  Ernst generating algorithm fails to add rotation to BBMB metric, difficulties arise also in the slow rotating approximation \cite{hideki}. The rotating metric of \cite{andres-hideki} does not have a proper mass term.
\item[$viii)$] Higher dimensional flavours of BBMB black hole are not known.
\end{itemize}

Thus, now, our purpose is to explore the possibility of a solution which is able to overcome these difficulties, or at least some. We restrict our research inside the most generic stationary axisymmetric Petrov type D class of metrics, which can be cast  in the Plebanski-Demianski form. Recently this issue, in presence of a scalar field coupling, was discussed in \cite{andres}. To begin with, we will focus on the the conformal coupling for the scalar field without the cosmological constant. \\
The most general non-stealt solution \footnote{For some values of the parameters $f_i$ there exist a matter configuration such that $T_{\mu\nu}=0$, although the fields $A_\mu$ and $\Psi$ are not null, so the matter does not have back reaction with the background spacetime.}, of the above form, admitting electromagnetic and  NUT charges, acceleration and in particular a standard mass\footnote{In the notation of \cite{andres} the mass term is related to odd powers of the $F(\xi)$ function.} and rotation terms, of the Kerr type in the limit of vanishing acceleration, requires a constant scalar field:
\bea \label{gen-metric-xy}
   \hspace{-0.7cm}  ds^2 &=& \frac{1}{(y-x)^2} \left[ \frac{F(y)(dt-x^2 d\varphi)^2}{1+x^2y^2} - \frac{1+x^2y^2}{F(y)} dy^2 + \frac{1+x^2y^2}{F(x)} dx^2 + \frac{F(x)(y^2dt+ d\varphi)^2}{1+x^2y^2}   \right]       \\
         \hspace{-0.7cm} F(\xi) &=& \sum_{i=0}^4 f_i \x^i \qquad \ \ \qquad \  A = \frac{ey(dt-x^2d\varphi)}{1+x^2y^2}  \  \qquad \ \ \qquad  \Psi =\pm\sqrt{\frac{6}{\k}} \ \sqrt{1+\frac{e^2}{f_0+f_4}}      
\eea
Constant conformally coupled scalar black hole metrics are not a novelty, some static solutions were already discussed in \cite{mtz} and \cite{dotti} for a slightly different theory including the cosmological constant (and an extra conformal $\Psi^4$ potential term in the action, usually associated with the presence of the cosmological constant).  \\
Even though the scalar field is constant it contributes non-trivially to the equations of motion (\ref{min-field-eq}). In fact, for a constant scalar field $\Psi_0$, from (\ref{min-field-eq}) we have:
\beq
         (1-\frac{\k}{6}\Psi_0^2) G_{\m\n}= \kappa T^{(EM)}_{\m\n}     
\eeq
Hence ( for $\Psi_0 \ne \pm \sqrt{6/\k}$) the presence of a constant conformally coupled scalar field has the property of rescaling the effective Newton coupling constant, thus rescaling the relative values of the electromagnetic charges. We will see hereinafter how the possibility of an arbitrary rescaling of the coupling constant, depending on the strength of the scalar field, has non-trivial physical effects. The basic difference with respect to the case with the cosmological constant (\cite{mtz} and \cite{dotti}) is that the value of the scalar field is not constrained by the coupling constants, as can be seen from (\ref{scalar-constr}).\\
 When the electromagnetic charges are vanishing a new branch of solutions is allowed with $\Psi_0 = \pm \sqrt{6/\k}$, whose supporting space-times do not have to be Einstein manifold, but they have to obey to the weaker condition coming from the scalar field equation (\ref{scalar-eq}): they simply are Ricci flat. It is possible to smoothly join these two branches in a unique family of metrics. To better clarify this point let's consider a specialisation of (\ref{gen-metric-xy}) without the NUT term and in spherical coordinates ($r=-1/Ay , \ \cos\t=x$):
\beq \label{hairy-acc}
 ds^2 = \frac{ \displaystyle \left[ \frac{r^4 G(r)}{\rho^2} (dt + a\sin^2\t d\varphi)^2 -\frac{\r^2}{r^4 G(r)} dr^2+ \frac{\r^2 \sin^2\t}{G(\t)} d\t^2 + \frac{G(\t)}{\rho^2} \left( a dt + (r^2+a^2) d\varphi \right)^2 \right] }{(1+Ar\cos\t)^2} 
\eeq
where
\bea 
         G(\xi) &=& (1-\xi^2)(1+r_+ A\xi)(1+r_- A\xi) \quad , \qquad  \xi=\{y=-1/Ar,\ x=\cos\t\}  \\
          \mathcal{A} &=& \frac{-erdt-are\sin^2\t d\varphi}{r^2+a^2 \cos^2 \t} \quad ,\\  \label{const-amu}
          \Psi &=& \pm \sqrt{\frac{6}{\k}} \sqrt{\frac{s}{s+e^2}} \quad ,  \label{constant-scalar}        \\
          r_\pm &=& m \pm \sqrt{m^2 - a^2 -e^2 - s} \quad , \label{horizons} \\
          \r &=& r^2 + a^2 \cos^2 \t   \quad ,  \label{rho}
\eea
$r_\pm$ points the positions of the inner and outer horizons, while the accelerating horizon is located at $\xi^2=1$, that is $r_A=\pm A^{-1}$.
This metric clearly describes an accelerating Kerr-Newman black hole dressed with a conformally coupled scalar hair, which is represented by the continuous parameter $s$. In fact, when the scalar parameter $s$ goes to zero, the standard accelerating Kerr-Newman black hole \cite{interpret-c} is recovered.
All sub-hierarchy of black holes until the Schwarzschild can be obtained switching on and off the parameters ($A,a,m,e,s$). In this sense the hair can be classified as primary hair,  contrary to the BBMB case.  \\
No hair theorems \cite{winstanley} are avoided because the scalar field is often assumed to vanish asymptotically or because it is not possible to connect this family of black holes with the Einstein frame by a conformal transformation, in the case of null electromagnetic charge.\\
The spacetime (\ref{hairy-acc})-(\ref{rho}) admits a further straightforward generalisation, in case of  cosmological constant (see appendix \ref{with-lambda}).\\

\subsection{Scalar hairy Reissner-Nordstrom black hole}

To have a clearer picture of the space-time described by the metric (\ref{hairy-acc}), let us consider a simpler case. When the rotation $a$ and acceleration $A$ parameters are null in (\ref{hairy-acc}) we remain with the static Reissner-Nordstrom black hole enriched by the scalar hair $s$
\beq \label{hairy-RN}
         ds^2 = - \left( 1- \frac{2m}{r} + \frac{e^2+s}{r^2}\right) dt^2 + \left(1- \frac{2m}{r} + \frac{e^2+s}{r^2} \right)^{-1} dr^2 +   r^2 d\t^2 + r^2 \sin^2 \t d\varphi^2 \quad .
\eeq
The scalar field remains the same as eq. (\ref{constant-scalar}) while the electromagnetic potential reduces to the standard RN one: $A_t=-e/r$. The total energy momentum tensor
\beq
           T^\m_{\ \n } = \frac{e^2+s}{r^4} \textrm{diag} (-1,-1,1,1)
\eeq
 satisfies both dominant and strong energy conditions whenever $s \geq - e^2$. Therefore, without violating these overall energy conditions\footnote{Anyway note that, when there is no electromagnetic field, the strong energy condition for the scalar field requires the positivity of $s$.}, it is even possible to erase the contribution of the electromagnetic field by means of the constant scalar field, just setting $s=-e^2$, hence recovering the Schwarzschild spacetime, but in this borderline case the scalar field become divergent. \\
Following the Ernst magnetising method for the accelerating version of this metric it is possible to remove the conical singularity without constraining any of the physical parameters $e,B,m,A$, because of the presence of the scalar parameter $s$. Furthermore this accelerating solution has not the curvature singularity of the BBMB C-metric, thus it is suitable to describe pair production of a scalar hairy black hole in the presence of an external magnetic field; these points are addressed in section \ref{app-acc-cost-bh}. While in appendix \ref{mcshcbh} a stationary, not accelerating, hairy Reissner-Nordstrom (RN) solution in an external magnetic field is generated from (\ref{hairy-RN}).   \\
It is evident by the similarities to the RN metric that the spacetime (\ref{hairy-RN}) has the same causal structure of the static charged black hole. The only difference now is that the position of the horizons is shifted by the presence of the scalar field constant parameter $s$, as can be seen from (\ref{horizons}), setting the rotation parameter $a=0$. The electric charge remains the same of the RN spacetime:
\beq
     Q = \frac{1}{4\pi} \int * F = e
\eeq
On the other hand, from a thermodynamical point of view, there are some dissimilarities with respect to the RN black hole, for instance about local stability; this point is addressed in the next section \ref{termo-hairy-rn}.  \\

\subsection{Thermodynamics of constant scalar hairy black hole}
\label{termo-hairy-rn}

To analyse the thermodynamics of the charged black hole with a conformally coupled constant scalar field (\ref{hairy-RN}) we will use the Euclidean method, as done in \cite{stafo}. The partition function for a thermodynamical ensemble is identified, around the Euclidean continuation of the classical solution, with the Euclidean path integral in the saddle point approximation \cite{gibbons-hawking}. Thus, first of all, we consider a minisuperspace of static Euclidean metrics given by:
\beq
     ds^2 = N(r)^2 f(r)^2 d\tau^2 + f(r)^{-2}dr^2 + r^2 d\Om^2 \ \ , 
\eeq
where the immaginary time $\tau$, obtained by a wick rotation $t \rightarrow i \tau$, has period $\b$, the inverse of the temperature $T$. It is obtained requiring regularity (no conical singularities in the ($\tau,r$) section) on the horizon:
\beq
      T = \frac{1}{\b} = \frac{N(r)}{4\pi} \left. \frac{d}{dr}  f(r)^2 \right|_{r_+} = \frac{r_+ - r_-}{4\pi r^2_+} \ \ .
\eeq
If the scalar field $\Psi(r)$ and the electromagnetic potential $A_\m(r)$ are considered to depend at most on the radial coordinate $r$, then the reduced euclidean action $\mathcal{I}$ becomes\footnote{Note that there is a sign discrepancy  with respect to \cite{stafo} because there the base manifold is hyperbolic.}  
\beq
       \mathcal{I} = \b \int_ {r_+}^\infty \left[ N(r) \mathcal{H}(r) + A_t \left(\frac{r^2}{N(r)} A'_t(r)\right)' \right] dr  \ + \ \mathcal{B} \quad ,
\eeq
where $\mathcal{B}$ is the surface term and the prime denotes the $d/dr$ derivative. The reduced Hamiltonian is given by
%\beq
%      \mathcal{H} = \frac{r^2}{2G} \left\lbrace \k \left[ \frac{f^2(\Psi')^2}{6}  - \frac{\Psi \Psi'}{6} \left( (f^2)' + \frac{4f^2}{r} \right)  - \frac{\Psi f^2 \Psi''}{3} \right] + \left( 1- \frac{\k\Psi^2}{6} \right) \left[ \frac{(f^2)'}{r} - \frac{1-f^2}{r^2} \right] + G\frac{(A_t')^2}{N} \right\rbrace \nn 
%\eeq
\beq
      \mathcal{H} = \frac{r^2}{2G} \left\lbrace \frac{\k}{6} \left[ f^2(\Psi')^2 - \Psi \Psi' \left( (f^2)' + \frac{4f^2}{r} \right)  - 2\Psi f^2 \Psi''\right] + \left( 1- \frac{\k\Psi^2}{6} \right) \left[ \frac{(f^2)'}{r} - \frac{1-f^2}{r^2} \right] + G\frac{(A_t')^2}{N} \right\rbrace \nn 
\eeq

In the grand canonical ensemble the variation of the action is implemented keeping the temperature fixed and the "injection voltage energy" $\Phi=A_t(\infty) - A_t(r_+)$. For the Euclidean solution under consideration $\Psi\propto$ constant, $N=1$, $\mathcal{H}=0$ and $(r^2 A_t')'=0$, so the variation of the action evaluated on the classical solution is just given by the variation of the boundary term $\d \mathcal{B}$.
\bea
     \d \mathcal{B} &=& - \frac{\b}{2G} \left[ r \left(1 - \frac{\k}{6} \Psi^2 \right) \d f^2 + 2G A_t \d (r^2 A'_t)  \right]^\infty_{r_+} \\
      \label{dB}    &=&  \left( \frac{e^2}{e^2 + s} \right) \frac{1}{G} \left(\b \d m - \frac{4 \pi r_+ \d r_+}{2} \right) - \b  \Phi \d e    \ \ ,
\eea
where the following boundary variations of the fields at infinity and at the horizon $r_+$ were used:
\bea
     \d(r^2 A'_t) \left. \right|_\infty &=& \d(r^2 A'_t)\left. \right|_{r_+} =  \d e \ \ , \\
     \d \Psi \left.   \right|_\infty  &=& \d \Psi \left.  \right|_{r_+} , \\
     \d f^2 \left. \right|_\infty &=&  - \frac{2}{r} \d m + O(r^{-2}) \ \ , \\
     \d f^2 \left. \right|_{r_+} &=&  -  (f^2)' \left. \right|_{r_+} \ \ .
\eea
Then defining an "effective Newton constant" $\tilde{G}$ as $ \tilde{G}^{-1} = G^{-1} e^2/(e^2 +s)  $ and integrating (\ref{dB}) we obtain the finite Euclidean action, up to an arbitrary additive constant:
\beq
      \mathcal{I} = \mathcal{B}(\infty) - \mathcal{B}(r_+) = \frac{\b}{\tilde{G}} m - \frac{A_+}{4\tilde{G}} - \b \Phi e \ \ .
\eeq
In the grand canonical ensemble the Euclidean action is related (in unit where Planck and Boltzmann constants are $\hbar=\k_B=1$) to the free energy by $\mathcal{F} = \b \mathcal{I}$. Thus the mass $M$, electric charge $Q$ and entropy $S$ are obtained by the usual thermodynamical relations:
\bea
      M &=&  \p_\b \mathcal{I} - \b^{-1} \Phi \p_\Phi \mathcal{I} = \frac{m}{\tilde{G}} \ \ \ , \\
      Q &=&  -\b^{-1} \p_\Phi \mathcal{I} = \ e  \ \ \ ,  \\
      S &=&  \b \p_\b \mathcal{I}  - \mathcal{I} =  \frac{A_+}{4\tilde{G}} 
\eea 
The first law of black hole thermodynamics is satisfied only using the effective Newton constant $\tilde{G}$, this is a typical feature of non minimal coupling of the scalar field \cite{stafo}. In the range of values of $s$ respecting the dominant and strong energy conditions the entropy remains positive.\\
While, when the scalar field is vanishing, for $s=0$, the standard results for Reissner-Nordstrom black hole are retrieved. It is interesting to observe that for the uncharged case ($e=0$) the total mass $M$ and the entropy $S$ become void.  Thus the scalar hair can be considered to be primary since it does not depend on the presence of the electric charge; anyway some physical spacetime properties are better behaved for $e\neq 0$.\\ 
A natural question is now if the charged constant hairy black hole (\ref{hairy-RN}) may decay into the Reissner-Nordstrom one, which is also a solution of the same action principle with a null scalar field, for a fixed temperature and electromagnetic potential injection. Evaluating the euclidean action, for fixed $\b$ and $\Phi$, for both RN and  (\ref{hairy-RN}) spacetimes there is not a stable thermodynamical favoured configuration, but it is possible to find numerically a critical point beyond which phase transitions can occur, for a certain range of parameters, that do not violate the strong and dominant energy conditions. \\
The local thermal stability with respect to the temperature fluctuation or electromagnetic fluctuation can be inferred by the analysis of the heat capacity at constant electric potential $C_\Phi$ and electrical permittivity at constant temperature $\e_T$ respectively, as done for the grand canonical ensemble  in \cite{Braden}
\beq \label{cphi}
          C_\Phi := T\left( \frac{\p S}{\p T} \right)_\Phi =  T \left(\frac{\p T}{\p r_+} \right)^{-1}_\Phi \left( \frac{\p S}{\p r_+} \right)_\Phi   =   -\frac{2 \pi r_+}{\tilde{G}} \ \frac{r^2_+ -e^2 -s}{r^2_+ -e^2 -3s} \ \frac{e^2+2s}{e^2+s}
\eeq
The local thermodynamical stability is given by the positivity of the heat capacity, thus according to (\ref{cphi}) and (\ref{horizons}) in this case the presence of the scalar field improves the local stability since there is a parametric window for which $C_\Phi\geq 0 $:
$$   \frac{r_+^2+e^2}{3} \leq s \leq m^2-e^2 \ \ .$$
The electrical permittivity is defined as
\beq \label{et}
           \e_T := \left( \frac{\p Q}{\p \Phi}\right)_T =  \left(\frac{\p \Phi}{\p r_+} \right)^{-1}_T \left( \frac{\p Q}{\p r_+} \right)_T \quad .
\eeq
But since the charge Q  only has dependence on terms of the potential at constant horizon we have to decompose each factor in the previous equation as
\bea
\left( \frac{\partial Q}{\partial r_+}\right) _{T} &=&- \left( \frac{\partial T}{\partial Q}\right) _{r_+}^{-1}\left( \frac{\partial T}{\partial r_+}\right) _{Q} \ , \\
\left( \frac{\partial \Phi }{\partial r_{+}}\right) _{T} &=& - \left( \frac{\partial T}{\partial \Phi }\right) _{r_+}^{-1}\left( \frac{\partial T}{\partial r_{+}}\right) _{\Phi } \ \ .
\eea
So electrical permittivity (\ref{et}) becomes 
\beq
          \e_T = \left(\frac{\p \Phi}{\p e} \right)^{-1}_{r_+}  \left(\frac{\p T}{\p r_+} \right)^{-1}_\Phi \left(\frac{\p T}{\p r_+} \right)_Q = r_+ \frac{r_+^2-3e^2-3s}{r_+^2-e^2-3s} \ .
\eeq
Therefore even from just a naive\footnote{Others thermodynalmical settings may be considered, even including an extra chemical potential for the scalar field.} thermodynamical study, we can see how the presence of the scalar field affects the local thermal stability of the solution (\ref{hairy-RN}) with respect to the Reissner-Nordstrom black hole for $s=0$.\\
In the next section we will present another application for which the presence of the scalar parameter $s$ has non-trivial physical consequences. \\

 \subsection{Magnetised accelerating constant scalar hairy charged black hole pair}\label{app-acc-cost-bh}

It could be of some interest to consider the magnetised version of the constant hairy charged and accelerating black hole, described by the metric (\ref{hairy-acc}), fixing, for simplicity, the rotation parameter $a=0$. This because it possess an extra free scalar parameter $s$, with respect to the no hairy one ($s=0$), which allows us to achieve a regular equilibrium solution (with no conical singularity) without imposing any constraints on the mass $m$, charge $g$, external magnetic field $B$ and acceleration $A$ parameters, as in the hairless case. \\
To keep the system as simple as possible we make use of the four dimensional electromagnetic duality, in the metric (\ref{hairy-acc}) with $a=0$, to obtain, as a seed, an intrinsic magnetically charged black hole instead of an electrically charged one:
\beq \label{hairy-pair}
      ds^2 =  \frac{1}{(1+Ar\cos\t)^2} \left[ - \textrm{g}(r) dt^2 - \frac{dr^2}{\textrm{g}(r)} + \frac{r^2 d \t^2}{p(\t)} + r^2 p(\t) \sin^2\t d\varphi^2 \right] \ , 
\eeq
where
\bea
         \textrm{g}(r) &=& (1-A^2 r^2) \left( 1 - \frac{2m}{r} + \frac{g^2 +s}{r^2} \right)  \qquad \ \ , \quad   \qquad  \Psi = \sqrt{\frac{6}{\k}}  \sqrt{\frac{s}{s+g^2}} \\ 
          p(\t) &=& 1 + 2 m A \cos \t + A^2 \cos^2\t (g^2+s) \qquad , \qquad  \   A_\varphi = -g \cos \t \ \ \ . 
\eea
This simplifies the discussion, because the solution after magnetisation remains static. Otherwise, to guarantee staticity, we might have considered alternatively the intrinsic electric charge, but then, in that case,  we should have embeded it in an external electric field. From a mathematical point of view this feature is portrayed by the fact that the Ernst potentials remain real (in the alternative case of an intrinsic electrically charged black hole in an external magnetic field, the electromagnetic Ernst potential $\mathbf{\Phi}$ remains purely imaginary, conversely the rotation is generated by fully complex potentials).  This point is addressed in appendix \ref{mcshcbh} in the case of null acceleration).  \\ 
Using the results of \cite{marcoa}, it is possible to write the Harrison magnetising transformation for this class of static magnetically charged spacetimes, directly in the Jordan frame:
\bea \label{harry-trans-magn}
             f &=&  \frac{f_0}{\L^2} \qquad \quad , \qquad \qquad \mathbf{\Phi} = \frac{\mathbf{\Phi}_0 + \frac{B}{2} \left[ \left( 1 -\frac{k}{6} \Psi^2 \right) f_0 - \mathbf{\Phi}_0^2 \right] }{\L}  \\
     &\textrm{where}&   \qquad \qquad \L = 1 - B \mathbf{\Phi}_0 - \frac{B^2}{4} \left[ \left( 1 -\frac{k}{6} \Psi^2 \right) f_0 - \mathbf{\Phi}_0^2 \right]
\eea
These are a generalisation of (\ref{harrison-jordan}) and (\ref{harry-transf}) in case of non vanishing intrinsic magnetic charge $g$, simpler expressed in terms of $f$ and $\mathbf{\Phi}$. \\
From the comparison with the Weyl metric (\ref{weyl-metric}) we can identify the needed seed functions
\beq
      f_0 = -  \frac{p(\theta)~ r^2 \sin^2{\theta}}{\left(1+A r\cos{\theta}\right)^2}   \qquad \qquad  \textrm{and} \qquad \qquad \mathbf{\Phi}_0 = - g \cos \t ~.
\eeq
Then after the action of the Harrison transform, according to  (\ref{harry-trans-magn}), we get the magnetised ones. Plugging these latter again in the Weyl metric (\ref{weyl-metric}) we obtain the magnetised version of (\ref{hairy-pair}):
\beq \label{magn-magn}
 ds^2 =  \frac{1}{(1+Ar\cos\t)^2} \left\lbrace \L^2 \left[ - \textrm{g}(r) dt^2 - \frac{dr^2}{\textrm{g}(r)} + \frac{r^2 d \t^2}{p(\t)} \right] + \frac{r^2 p(\t) \sin^2\t}{\L^2} d\varphi^2  \right\rbrace \ , 
\eeq
where the functions g$(r)$, $p(\t)$ and $\Psi$  remain the same as the non magnetised solution, while the electromagnetic potential
$$ A_\varphi = - \frac{ \displaystyle g \cos \t + \frac{B}{2} \left[ \frac{g^2}{g^2+s}\frac{p(\theta)~ r^2 \sin^2{\theta}}{\left(1+A r\cos{\theta}\right)^2} + g^2 \cos^2 \t \right]}{ \displaystyle 1 + B g \cos \t + \frac{B^2}{4} \left[  \frac{g^2}{g^2+s} \frac{p(\theta)~ r^2 \sin^2{\theta}}{\left(1+A r\cos{\theta}\right)^2} + g^2 \cos^2 \t \right] } , $$ 
which supports (\ref{magn-magn}), include both the intrinsic magnetic charge of the black hole and an external Melvin-like magnetic field. The metric (\ref{magn-magn}) describes a pair of accelerating magnetically charged black holes in presence of an external magnetic field and a conformally coupled scalar field. When the scalar field is null, that is $s=0$, we recover the Ernst solution \cite{ernst-remove}. \\
As usual, accelerating metrics (\ref{magn-magn}) posses a couple of conical singularities on the poles, one (let's say around $\t=0$) is always easy to remove, following the same analysis of section \ref{conformal-C-metric}, by rescaling the angular coordinate $\varphi$, such that:
\beq
      \varphi \longrightarrow \phi = \frac{1+2mA+A^2(g^2+s)}{\left( 1+ \displaystyle \frac{ Bg}{2}\right )^4} \varphi \ ,
\eeq
while the second singularity can be removed thanks to a constraint relation between the physical parameters $m,g,B,A,s$. An interesting feature of the conformally coupled constant scalar field solution is that it introduces a new parameter $s$ with respect to the Reissner-Nordstrom spacetime, which, when expressed in terms of mass, acceleration and intrinsic magnetic charge, allows us to remove the second conical singularity for $\t=\pi$, without fine tuning  these latter charges as in the Ernst solution \cite{ernst-remove}:
\beq
         s = \frac{m \left( 1+\frac{2}{3} g^2 B^2 + g^4 B^4  \right)}{A \left( gB+ \displaystyle \frac{B^3g^3}{4} \right)} - \frac{1}{A^2} -g^2 \quad .
\eeq  
Therefore, even though the effect of the constant scalar field is not dynamical and it reduces just to an effective rescaling of the Newton constant, it opens  to the possibility of modelling less constrained magnetised charged black holes with respect to the null scalar field case. This feature has the effect that more general black holes in the pair creation process (in the spirit of \cite{hawking}, \cite{kastor}) can be admissible in a strong magnetic background, and also the pair creation rate\footnote{The pair creation rate is obtained (see \cite{hawking}, \cite{kastor} for details) as the difference of the action evaluated on the lukewarm instanton and the action evaluated on the Melvin magnetic background. The lukewarm instanton can be produced as the Wick rotated $t \rightarrow i \tau$ metric (\ref{magn-magn}) regularised from conical singularities, in the Euclidian time, such that the temperature of the event and acceleration horizons coincides.}  is affected by the extra parameter $s$. This is so because pair creation probability depends on the position of the roots of the  g$(r)$, which is modified with respect to RN spacetime whenever $s\ne 0$.  \\
Another possibility to regularise the spacetime (\ref{hairy-pair}), without resorting to an external electromagnetic field, consists in directly fine-tuning the constant scalar field and rescaling the azimuthal coordinate in order to cancel the angular singularity of the C-metric. But this can be done in a slightly different, not equivalent, radial coordinate, the one used for C-metric before \cite{hong-teo}. 
\\

%%%%%%%%%%%%%%%%%%%%%%%%%%%%%%%%%%%%%%%%%%%%%%%%%%%%%%%%%%%%%%%%%%%%%%%%%%%%%%%%%%%%%%%%%%%%%

\section{Comments and Conclusions}

In this paper the Ernst solution generating technique, in the context of Einstein-Maxwell gravity conformally coupled to a scalar field, is applied to a C-metric solution, which describes a couple of accelerating BBMB black holes.  Through a Harrison transformation we manage to embed the BBMB C-metric into an external magnetic field. The resulting solution shows more regularity than the no accelerating one, but still it presents a curvature singularity on a pole of the event horizon, due to a divergence of the scalar field at that point.  Thanks to this regularity enhancement we are able to build a fully regular black hole metric by a cut and paste procedure. The price to pay was the introduction of extra matter on the thin shell gluing surface. \\  
Therefore a better behaved seed solution, that is able to overcome several disadvantages of the BBMB spacetime, is considered for the theory under consideration. The requirements of a proper mass term and rotation constrain the scalar field to be constant, at least in the realm of the Plebanski-Demianski spacetimes\footnote{Therefore an eventual stationary generalisation of the BBMB black hole have to be searched for outside the Plebanski-Demianski ansatz.}.  In that case it is possible to write a regular black hole family of solutions, comprising the Kerr black hole and featuring acceleration, mass, rotation, intrinsic electromagnetic charge and an extra scalar parameter. The thermodynamical properties of a simple black hole of this family (without acceleration and rotation) are studied and compared to the vanishing scalar field case, the Reissner-Nordstrom black hole.  By a Harrison transformation we were able to embed some black holes of this family in an external magnetic field.  It is interesting to note that the presence of the scalar field introduces an extra parameter $s$, which can be tuned (in terms of the other physical parameters) to cancel the string, encoded in the conical singularity, that is pulling the two black holes. This is the main difference compared to the case without the scalar field $s=0$. A completely regular balanced solution can be obtained without constraining between themselves the mass, intrinsic charge, acceleration and external magnetic strength.  Possibly this is an astrophysically observable feature for the black hole family considered. Of course another possible observable property is the correction to the standard Newton law due to the presence of the scalar field, which, for example, can be tested in galaxies rotation curves. An upper limit constraint to the value of the scalar parameter $s$ can also be found from solar system physics.  
%A non null constant scalar field can also be fine-tuned to remove the string/strut between the two black hole of the hairy C-metric, thus regularising it without an extra magnetic field.\\ 
\\
It may also be interesting, for a future perspective, to study if this constant scalar field gives some contribution on cosmological level, in particular concerning the open problems of amount of dark energy and dark matter.\\ 
Eventually people interested in higher dimensions may find the four dimensional C-metrics presented in this paper of some utility in building novel, topological non trivial solutions in five dimensions.\\

\section*{Acknowledgements}
\small I would like to thank Theodoros Kolyvaris,  Hideki Maeda, Cristi\'{a}n Mart\'{i}nez, Tim Taves, Ricardo Troncoso and Jorge Zanelli for fruitful discussions. 
\small This work has been funded by the Fondecyt grant 3120236. The Centro de Estudios Cient\'{\i}ficos (CECs) is funded by the Chilean Government through the Centers of Excellence Base Financing Program of Conicyt.
\normalsize
\\

%%%%%%%%%%%%%%%%%%%%%%%%%%%%%%%%%%%%%%%%%%%%%%%%%%%%%%%%%%%%%%%%%%%%%%%%%%%%%%%%%%%%%%%%%%%%%%%%%

\appendix

\section{Magnetised stationary charged constant scalar hairy black hole}
\label{mcshcbh}

When the constant hairy Reissner-Nordstrom spacetime (\ref{hairy-RN}) is embedded in an external magnetic field, the solution acquires angular momentum for the reason commented in section \ref{remove-cony}. In order to proceed in the magnetisation process the metric (\ref{hairy-RN}),  as explained in \cite{marcoa}, have first to be lifted in the Einstein frame:
\beq
      ds^2 = \frac{e^2}{e^2+s} \left[- \left( 1- \frac{2m}{r} + \frac{e^2+s}{r^2}\right) dt^2 + \left(1- \frac{2m}{r} + \frac{e^2+s}{r^2} \right)^{-1} dr^2 +   r^2 d\t^2 + r^2 \sin^2 \t d\varphi^2 \right] . \nn
\eeq
then comparing with the Weyl metric (\ref{weyl-metric}) is possible to find the seed Ernst complex potentials:
\beq
        \Er_0 = - \frac{e^2}{e^2+s} r^2 \sin^2 \t - e^2 \cos^2 \t  \qquad \qquad \mathbf{\Phi}_0= -i e \cos \t 
\eeq
We recall the definitions of Ernst potentials, according to the notation of \cite{marcoa}:
\bea \label{ernst-pot-def}
        \mathbf{\Phi} &:=& A_\varphi + i \tilde{A}_t  \qquad , \qquad \qquad     \Er := f - |\mathbf{\Phi} \mathbf{\Phi}^*| + i h  \quad , \\
          \overrightarrow{\nabla} \tilde{A}_t &:=& - f R^{-1} \overrightarrow{e}_\varphi \times (\overrightarrow{\nabla} A_t + \omega \overrightarrow{\nabla} A_\varphi ) \qquad , \qquad \\
            \overrightarrow{\nabla} h &:=& - f^2 R^{-1} \overrightarrow{e}_\varphi \times \overrightarrow{\nabla} \omega - 2 \ \textrm{Im} (\mathbf{\Phi}^*\overrightarrow{\nabla} \mathbf{\Phi} )  \qquad . \label{ernst-pot-def2}
\eea
Then we act on them by a Harrison transformation to get the magnetised ones:
\bea
      \Er &=& \frac{\Er_0}{1-B\Phi_0-\frac{B^2}{4} \Er_0} = \frac{ - \frac{e^2}{e^2+s} r^2 \sin^2 \t - e^2 \cos^2 \t}{1 + ieB\cos \t + \frac{B^2}{4}\left( + \frac{e^2}{e^2+s} r^2 \sin^2 \t + e^2 \cos^2 \t \right)}  \\
      \mathbf{\Phi} &=& \frac{\Phi_0 + \frac{B}{2}\Er_0}{1-B\Phi_0-\frac{B^2}{4} \Er_0} = \frac{ -i e \cos \t - \left( \frac{e^2}{e^2+s} r^2 \sin^2 \t + e^2 \cos^2 \t \right)}{1 + ieB\cos \t + \frac{B^2}{4}\left(  \frac{e^2}{e^2+s} r^2 \sin^2 \t + e^2 \cos^2 \t \right)}~.
\eea
Thus the magnetised  intrinsically charged metric, once uplifted again in the Jordan frame, becomes
\bea \label{hairy-mag-rn}
            ds^2 &=&  | \L |^2 \left[- \left( 1- \frac{2m}{r} + \frac{e^2+s}{r^2}\right) dt^2 + \left(1- \frac{2m}{r} + \frac{e^2+s}{r^2} \right)^{-1} dr^2 +   r^2 d\t^2  \right]   + \nn \\
            & & \qquad \ \ + \  r^2  \frac{\sin^2 \t}{|\L|^2} \Big[d\varphi - \omega(r,\t)dt \Big]^2 \ .
\eea 
For the Harrison transformation under consideration, $\omega(r,\t)$ can be found, using (\ref{ernst-pot-def2}), from the relation $ \overrightarrow{\nabla} \omega = \Lambda \Lambda^* \overrightarrow{\nabla} \om_0 - i \overrightarrow{e}_\varphi \times \frac{R}{f_0} (\Lambda^*\overrightarrow{\nabla}\Lambda-\Lambda\overrightarrow{\nabla}\Lambda^*)$:
\bea
         \omega(r,\t) &=& \frac{e^3 B^3}{e^2+s} \sin^2 \t \   \frac{r^2 -2mr +e^2 +s }{2r} + \frac{2Be}{r} - \frac{B^3e^3}{2r} - \frac{B^3 e^3}{e^2+s} r \\
         \L &=& 1 + ieB\cos \t + \frac{B^2}{4}\left( \frac{e^2}{e^2+s} r^2 \sin^2 \t + e^2 \cos^2 \t \right)
\eea
Finally, thanks to eqs. (\ref{ernst-pot-def})-(\ref{ernst-pot-def2}), the electromagnetic potential supporting this metric results
\bea
          A_t  &=& - \frac{3 e^3B^2}{4 (e^2+s)} \sin^² \t \left(  r - 2m + \frac{e^2+s}{r}  \right) + \frac{3 e^3B^2}{2 (e^2+s)} r +\frac{3 e^3B^2}{4 r} - \frac{e}{r} - \omega A_\varphi\\
          A_\varphi &=& \textrm{Re} \ \mathbf{\Phi} =  |\L|^{-2} \left[ -\frac{B}{2}  \left( \frac{e^2}{e^2+s} r^2 \sin^2 \t + 3 e^2 \cos^2 \t \right)  - \frac{B^2}{8} \left(  \frac{e^2}{e^2+s} r^2 \sin^2 \t + e^2 \cos^2 \t \right) \right]  \nn
\eea
The Reissner-Nordstrom spacetime in an external magnetic field \cite{ernst-magnetic} is a sub case of (\ref{hairy-mag-rn}) corresponding to $s=0$. While when $e=0$ the electromagnetic field becomes null, so in this case the Harrison transformation was not able to magnetise the metric (\ref{hairy-RN}).\\
The spacetime (\ref{hairy-mag-rn}) follows the usual characteristic of magnetised, stationary, charged black holes. Thus it presents a removable conical singularity on the poles, that can be erased by a redefinition of the azimuthal coordinate
\beq
\varphi \longrightarrow \phi = \frac{\varphi}{1+\frac{3}{2}e^2 B^2+ \frac{1}{16}e^4 B^4}   \label{deficit-angle}\  \ .
\eeq 
Therefore the period of the azimuthal coordinate become $\Delta \phi=2\pi / (1+\frac{3}{2}e^2 B^2+ \frac{1}{16}e^4 B^4)$.
Regarding the topological properties of the event horizon, let's consider a two-dimensional surface $\mathcal{\bar{S}}$ of constant time $\bar{t}$ and radius $\bar{r}$.  It easy to check, by the Gauss-Bonnet theorem, that the black hole has an event horizon of the same topology of the spherical unmagnetised one:
\beq
      \chi(\bar{\mathcal{S}})=\frac{1}{4\pi}  \int_{\bar{\mathcal{S}}} \sqrt{\bar{g}} \ \bar{\textrm{R}} \ d\t \ d\phi \   =  2 \quad . \nn  
\eeq
The area is smaller than the regular spherical one, because the deficit angle (\ref{deficit-angle})
$$ \mathcal{A} = \int_0^{\Delta\phi} d\phi \int_0^\pi d\t \sqrt{g_{\t\t}} \sqrt{g_{\phi\phi}} = \frac{4\pi r_+^2}{1+\frac{3}{2}e^2 B^2+ \frac{1}{16}e^4 B^4} \quad .$$
And the horizon has a prolate geometry, stretched in the direction of the magnetic field, as one can easily check from inspection of the polar circumference
%\beq
%     C_e =   \int_0^{2\pi}   d\t \sqrt{g_{\t\t}}  = ...
%\eeq
and equatorial circumference $C_e$. For instance the equatorial circumference is shrunk by a factor proportional to the deficit angle and a factor proportional to $\L(r_+,0)$:  
\beq
      C_ e=   \int_0^{\Delta \phi} d\phi   \sqrt{g_{\phi\phi}} = \frac{\Delta \phi \  r_+}{1+ \frac{B^2}{4} \frac{e^2 r^2_+}{e^2+s} }  \nn \ .  
\eeq
Thus the polar circumference is stretched along the $z$-axe.

\section{Accelerating, rotating, charged constant hairy black hole with cosmological constant}
\label{with-lambda}

A further generalisation of the metric (\ref{hairy-acc}), describing an accelerating, rotating and intrinsically charged black hole with a conformally coupled, constant scalar hair can be found when we are in presence of two additional terms in the action due to cosmological constant $\lambda$, and also due to an extra scalar conformally coupled potential $\a \Psi^4$. We present it here, but because the Harrison transformation in presence of  $\lambda$ is not known at the moment, it will not be possible to embed it in an external electromagnetic field. The equations of motion are modified with respect to the null cosmological ones (\ref{min-field-eq}), in fact the scalar energy momentum tensor $T^{(S)}_{\m\n}$ and scalar equation became:
\bea
       T^{(S)}_{\m\n} &=&  \p_\mu\Psi\p_\nu\Psi-\frac{1}{2}g_{\mu\nu} \p_\s\Psi\p^\s\Psi+\frac{1}{6}\left[g_{\mu\nu}\square-\nabla_\mu\nabla_\nu+G_{\mu\nu}\right]\Psi^2  - \a g_{\m\n} \Psi^4  \ \ ,   \\
        \Box \Psi &=& \frac{1}{6} \textrm{R} \Psi + 4 \a \Psi^3 \ \ .
\eea
The electromagnetic equations remain the same as (\ref{eltr-eq}), so the potential $A_\m$ also remains unchanged as in (\ref{const-amu}) (and also $\r$, $\Psi$), while the metric in presence of the cosmological constant become: 
\beq
  ds^2 = \frac{  \displaystyle \left[ -\frac{F(r)}{\rho^2} (dt + a\sin^2\t d\varphi)^2 +\frac{\r^2}{F(r)} dr^2+ \frac{\r^2 }{G(\t)} d\t^2 + \frac{G(\t)}{\rho^2} \sin^2\t  \left( a dt + (r^2+a^2) d\varphi \right)^2 \right] }{ \displaystyle (1+Ar\cos\t)^2}~,
\eeq
where
\bea
    \hspace{-0.0cm}        F(r) &=& (1-A^2r^2)\left[r^2-2mr+e^2+s+a^2 \left(1+\frac{\l}{3A^2}\right) \right] - \frac{\l}{3}\left( r^4 + \frac{a^2}{A^2} \right) \\
 \hspace{-0.0cm}          G(\t) &=& 1 + 2Am\cos\t + A^2 \cos^2 \t \left[ e^2 + s +a^2 \left(1+\frac{\l}{3A^2}\right)  \right] \\
   \hspace{-0.0cm}         \a &=& -\frac{\k \l}{36} \ \frac{s+e^2 }{s} \quad . \label{scalar-constr}
\eea
The causal structure is the same as the standard accelerating, charged and rotating C-metric (which can be obtained in the smooth $s \rightarrow 0$ limit). The basic difference with respect to this latter, apart from the fact that the horizons are shifted in
\beq
             r_\pm = m \pm \sqrt{m^2 - e^2 - s - a^2\left(1+ \frac{\l}{2 A^2} \right) }  \ \ ,
\eeq
is that the scalar hair parameter $s$ allows one to set the strength of the scalar field and thus to arbitrarily tune the value of the coupling constants. As we have seen in section \ref{app-acc-cost-bh} this feature can have relevant astrophysical consequences, at least in the balance between the string and external magnetic field strength of the magnetised C-metric.  The   vanishing cosmological constant limit is well defined and gives the solution (\ref{hairy-acc})-(\ref{rho}).

%%%%%%%%%%%%%%%%%%%%%%%%%%%%%%%%%%%%%%%%%%%%%%%%%%%%%%%%%%%%%%%%%%%%%%%%%%%%%%%%%%%%%%%%%%%%%%%%

\end{document}